# Questions and Issues arising at SCES '94

P. Coleman

*Serin Physics Laboratory, Rutgers University, PO Box 849, Piscataway NJ, 08855.*

## Abstract

I summarize some of the key questions to have emerged during the 1994 conference on "Strongly Correlated Electron Systems", held in Amsterdam, August 1994. Issues addressed include: Hunds rule interactions and how they renormalize; the Luttinger sum rule and metamagnetism; heavy fermion insulators, the nature of the charge gap, spectral weight transfer in the optical conductivity; non-Fermi liquid behavior in transition and heavy fermion metals; order parameter symmetry and the unusual nature of quasiparticle excitations in heavy fermion superconductors.

Few attending this conference could fail to be struck by the breadth of topics. As I worried about how I might summarize the theory, I was inspired by a quote from a popular nineteenth century theorist:

*Like all other arts, the Science of Deduction and Analysis is one which can only be acquired by long and patient study ... Before turning to those moral and mental aspects of the matter which present the greatest difficulties, let the inquirer begin by mastering more elementary problems.*

Sherlock Holmes, "A Study in Scarlet".

Here in Amsterdam, new ideas and results frequently confronted us with old issues once thought to be settled. This is really a commentary on some of the questions that arose.

One of the guiding concepts in heavy fermion (HF) physics is Landau's notion of "adiabaticity" : the simple, yet bold idea that despite the immensity of the bare interactions between f-electrons, normal state excitations may be mapped onto states of a non- interacting system by adiabatically "turning on" the interactions.[1,2] This conference returned to re-examine these assumptions from various angles: it was the first SCES conference with extensive theoretical and experimental discussions about non-Fermi liquid behavior. Worries about Luttinger's theorem, and the effectiveness of a renormalized band picture surfaced many times. In addition, there were various new ideas about the nature of HF antiferromagnetism and its interplay with HF superconductivity.

## 1. Single impurity and normal state physics

A third of the experimental results involved uranium HF compounds. Paradoxically, as emphasized by Allen,[3] we do not have a basic microscopic model for the normal state of these systems. Unlike cerium heavy fermion compounds, uranium HF ions contain two, or three f-electrons. This brings into play Hunds rule interactions whose effects on Fermi liquid behavior are essentially <u>uncharted</u>. In its simplest form, the Hunds rule interaction may be written as a ferromagnetic exchange between the total ($SU(2)$) angular momentum operator $\vec{\mathcal{J}}_f = f^\dagger_\alpha \vec{J}_{\alpha\beta} f_\beta$ of the f-electrons

$$H_{Hunds} = -\Gamma_H (\vec{\mathcal{J}}_f)^2 \qquad (1)$$

These interactions involve shape, rather than charge fluctuations and are unscreened from their ionic values in the bulk: in Uranium, $\Gamma_H \sim 1\ eV$, a vast scale that effectively locks the f-electrons into a single large moment.[4–6]

It is difficult to reconcile this view with the observation of f-bands, *unless* the Hunds rule coupling is somehow renormalized down to the the HF bandwidth. If this is the case, the renormalization process and the residual Hunds interactions at low temperature could play a key role in heavy fermion physics.[6] Surprisingly, this issue has not been addressed even in the simplest, Hunds-rule coupled Anderson impurity model:

$$H = \sum \epsilon_k c^\dagger_{km} c_{km} + \sum_{k,\,m} V[c^\dagger_{km} f_m + \mathrm{H.c.}] + E_f n_{fm} + U \sum_{m>m'} n_{fm} n_{fm'} - \Gamma_H (\vec{\mathcal{J}}_f)^2$$

where $m \in [-J, J]$ labels the azimuthal quantum number of the f-electron and conduction electron partial wave states with spin $J$.[7] Let me encapsulate these thoughts as a question:

• *In the degenerate Anderson impurity model with large Hunds rule coupling $\Gamma_H$, is there an analogue of the Kondo effect, setting an energy scale $T^*$ below which the Hunds-rule coupled f-moment decouples into independent f-electrons, characterized by a renormalized Friedel Anderson resonance and a weak residual Hunds-rule coupling?*

The basic approach required to answer this question is set out in the pioneering work of Nozières and Blandin,[4] and I suspect a strong coupling expansion of a Hunds-coupled Kondo model will confirm this renormalization process. Actually, uranium ions in HF superconductors exist in either magnetic Kramers $f^3$ (e.g. $UPt_3$) or singlet non-Kramers configurations ($f^2$) (e.g. $URu_2Si_2$ or $UPd_2Al_3$). A renormalization of Hunds rule coupling to weak coupling would restore magnetic properties to non-Kramers ions at temperatures $T < T^*$. This might account for the remarkable insensitivity of HF superconductivity to such radical differences in bare ionic configuration.

Shimizu et al.[10] presented a closely related piece of work. They considered a model of an $f^2$ atom with an *antiferromagnetic* Hunds coupling. These authors carried out a numerical renormalization group calculation to find that when the triplet-singlet splitting $\Delta$ exceeds the characteristic temperature $T^*$ of the spin fluctuations, the Friedel-Anderson resonance is lost. It would be interesting to know whether the same phenomenon occurs for a *ferromagnetic* Hunds interaction.

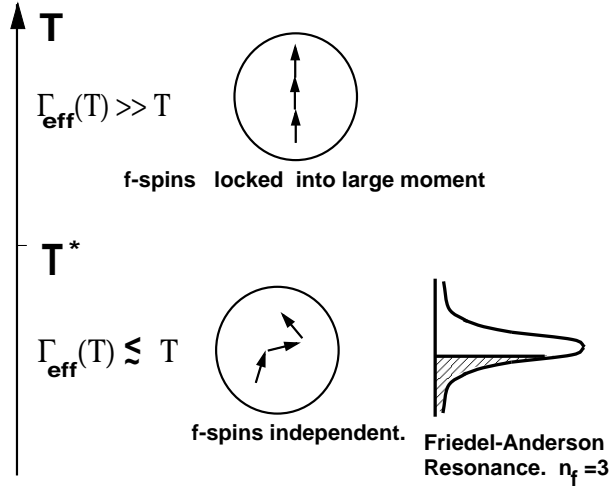

Fig. 1: **Renormalization of Hunds Interaction.** Illustrating hypothetical crossover from strong Hunds coupling ($T > T^*$) to weak Hunds coupling that is presumed to occur in uranium HF ions.

It is over a decade since Martin[8] emphasized that "adiabaticity" implies that the Fermi surface volume must count both conduction and the localized f-spins, according to Luttingers sum rule:

$$\frac{V_F}{(2\pi)^3} = n_c + n_f$$

Several papers considered situations where f-electrons start to recover local-moment properties, removing f-weight from beneath the Fermi surface ("$n_f \to 0$"). One feature of HF systems to which such considerations may apply is the metamagnetic transition. "Metamagnetism" refers to the sudden increase in magnetization that develops at a certain threshold applied field in many HF metals. Two experimental groups[11,12] reported large-scale trans-formations in the Fermi surface of $CeRu_2Si_2$ at its metamagnetic transition, which they interpret as a transition of f-electrons from valence to core-states. It is unclear whether this transition occurs via a rapidly developing polarization of heavy bands, or whether a "two fluid" picture of the f-electrons is required. In a similar context, Varma[13] pointed out the difficulty in understanding the transition from insulating to metallic $SmB_6$ under pressure, without invoking the notion that the f-electrons localize at high pressure. These two situations appear to pose the question:

• *Under what conditions does the "Luttinger" sum rule apply; can metamagnetic transitions be understood within a renormalized band picture, or is it necessary to invoke a "two-component" normal state whereby part of the f-electrons are localized, and part are delocalized?*

## 2. Kondo Insulators

"Kondo insulators" proved to be an issue of fervent discussion. Although simplistically, these systems can be regarded as highly renormalized band insulators,[14] where f-spins behave as valence electrons, the precise nature of the spin and charge gap, and its development at low temperatures is a subject of great concern.

A topic of particular note concerned the spectral weight transfer that accompanies development of the optical gap in Kondo insulators. The total integral of the optical conductivity determines the plasma frequency $\omega_p$, a quantity which depends only on the charge density.

$$\int \frac{d\omega}{\pi}\sigma(\omega) = \frac{\omega_p^2}{4\pi} \qquad (2)$$

In a band insulator, development of a gap in $\sigma(\omega)$ is expected to redistribute of spectral weight just above the gap. How does the optical spectral weight redistribute in a Kondo insulator? Schlesinger et al.[15] have reported FIR data on the Kondo insulators, $FeSi$, $Ce_3Pt_4Bi_3$ that indicate spectral weight from the gap is transferred to high energies, far in excess of the gap. At this meeting, the Zurich group[16] presented conflicting data where the missing spectral weight from the gap is recovered within energies of order $2-3$ times the gap.

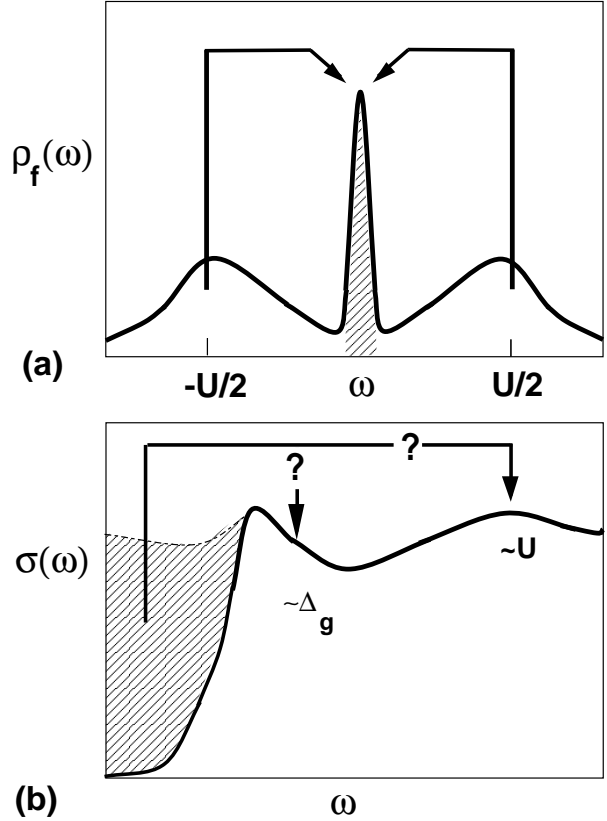

Fig. 2: **Spectral weight transfer.** Illustrating (a) large downwards spectral weight transfer in the f-spectral function of the symmetric impurity Anderson model, and (b) the question of whether optical spectral weight in the Kondo insulator is transferred to energies of order the interaction scale.

Large spectral-weight transfers are not unknown in the context of strongly correlated electron systems. In the single impurity Anderson model for example, the Kondo resonance in the f-spectral function is formed rapidly at temperatures comparable with the Kondo temperature $T_K$ by transfer of spectral weight from much higher energies of order $\pm U$ above and below the Fermi energy.(Fig. 2.) It would be intriguing if, in the Kondo insulators, we have an example where upward spectral weight transfer to high energy occurs as the temperature is lowered. This would would seem to be an ideal topic for the various large d approaches discussed here.[17-20]

A variety of numerical studies of Kondo insulator models were presented which demonstrate[21-23] that the charge and spin charge gap in these systems can differ markedly, especially in situations close to a magnetic instability, when the ratio of spin to charge gap appears to go to zero. Maekawa et al. presented results for a 2D Kondo insulator[23] where the excitations above the gap resembled excitations in a system with a doubled unit cell, suggesting incipient antiferromagnetic correlations. Experimentally, there is evidence for a suppression of spin gaps relative to charge gaps.[24] The Kondo insulators $CeNiSn$ and $CeRhSb$, with the smallest charge gaps appear to display gapless spin excitations, as evidenced by their $T^3$ NMR relaxation rate.[25] Kondo insulators appear to be on the brink of antiferromagnetism: an aspect that has not, as yet received much attention in theoretical models.[26,27]

Only one paper discussed the important effect of impurities in HF insulators. Schlottmann et al describe[28] the effects of vacancies on Kondo insulators. By removing an f-electron, a vacancy in a Kondo insulator creates a localized conduction electron bound state, or "Kondo hole". When these states overlap an impurity band develops, whose width they find should scale as the square-root of the vacancy concentration. Experimentally, no HF "insulator" is truly insulating, posing the question:

- *What level of impurity vacancies is required to drive Kondo insulators metallic?*

Presumably, the critical concentration $n_c$ is set by the typical size of the Kondo-hole wavefunction $l$, $n_c \sim l^{-3}$.

## 3. Non-Fermi Liquid behaviour

Motivated by a strong collective desire to elucidate the nature of the unusual normal state in cuprate superconductors, interest in extending the number of theoretical and experimental examples of "non Fermi Liquid" (NFL) properties has blossomed in recent years. In the home town of Heineken, a much appreciated non-Fermi Liquid, it was perhaps appropriate that we heard so many good theoretical and experimental talks on this topic.

Theoretically, non-Fermi liquid behaviour is expected to occur at a "quantum phase transition": a second-order phase transition at $T_c = 0$. At this conference there were several experimental groups that reported successful attempts to "create" and study non-Fermi liquid metals by fine tuning to a quantum critical point.[30-34] Millis described how close to a quantum phase transition,[29] interactions between electrons are mediated by the exchange of overdamped massless (magnetic) fluctuations with a Landau Ginzburg action of the form

$$S[\phi] = \sum_{\vec{q},\nu_n} \left( \left[ \frac{|\nu_n|}{\Gamma_{\vec{q}}} + 1 \right] cq^2 \right) |\phi_{\vec{q}\nu}|^2 + U\phi^4$$

These fluctuations are "dangerously irrelevent": dangerous, because they qualitatively modify the physics; irrelevant because the strong momentum dependence of their decay rate $\Gamma(q) \propto q^z$ raises raises the effective dimensionality of the theory above the upper critical dimension for $\phi^4$ field theory

$$d_{eff} = d_{spatial} + z > d_c = 4$$

so that predictions of the Hartree or "Gaussian" approximation are essentially exact.

McMullen described how these ideas can be experimentally tested[30] in itinerant weak ferromagnets $MnSi$ and $ZrZn_2$. These compounds can be tuned to a quantum phase transition by suppressing their Curie temperature under pressure. In $ZrZn_2$, where there are no complications with first-order transitions, the temperature dependence of the critical resistivity and the pressure dependence of the transition temperature beautifully agree with the the predictions of quantum critical phenomenon.

$$\frac{\rho(T)}{T^2} \propto T^{-1/3}, \qquad T_c \propto (p - p_c)^{3/4}$$

Von Löhneysen, Aronson and various other groups[31-34] presented convincing evidence for the development of non-Fermi liquid behavior at a quantum critical point in systems, such as $CeCu_{6-x}Au_x$, and $UCu_{5-x}Pd_x$ where chemical pressure from doping drives an an antiferromagnetic transition temperature to zero. Though scale-invariant behavior is observed in transport, thermodynamic and neutron data, a consistent explanation of the specific heat and susceptibility in terms of a single scaling form for the free energy

$$F = T^{1+d_{spatial}/z} f[H^2/T^{2\Delta}]$$

does not seem to be possible. Here it appears that the presence of disorder may change the theoretical picture significantly, raising the question

- *What are the effects of disorder on a quantum critical point?*

Though the "quantum critical point" provides a very useful example of non-Fermi liquid behavior, the direct applicability of this model to the normal state of cuprate superconductors is controversial. Two groups discussed the possibility of an antiferromagnetic quantum critical point as the origin of NFL behavior in the cuprate superconductors.[35,36] Here two difficulties seem to remain: (i) a $T^{3/2}$ temperature dependence of the resistivity occurs with the most natural choice $z = 2$, and (ii) the need to fine-tune the system to the special point where $T_c = 0$ to avoid a cross-over into a Fermi liquid state.

NFL behaviour can also systems occur in low-dimensional, or impurity systems. A topic of continued interest is the non-Fermi liquid behavior of the two-channel Kondo model.[5] Here, scattering off an unusual localized zero-mode with *fractional* entropy is responsible for the NFL behaviour. In a lovely paper, Suga et al. show[37] how this zero mode appears as a sharp delta function excitation in the spinon density of states. Let then pose:

- *What is the microscopic character of the local zero-mode responsible for NFL behavior in the two channel Kondo model, and are there conditions where such a mode would occur without fine-tuning the model?*

Suga et al. show that in the presence of an applied field, this mode couples to the Fermi sea, broadening into a Lorentzian, and leading to Fermi liquid behavior at a finite field.

## 4. Heavy Fermion Magnetism

A long-standing question in heavy fermion physics concerns the nature of the antiferromagnetism, specifically, the origin of ordered magnetic phases with tiny moments, and the co-existence of local moment antiferromagnetism and superconductivity. These issues are exemplified by $URu_2Si_2$, which undergoes a sharp mean-field transition at $17K$ with a large specific heat anomaly, yet develops only a tiny ordered moment $0.03\mu_B$. Three contrasting mechanisms of this ordering transition were discussed:

- Sikkemma et al.[38] examined the possibility that this phase is a nested antiferromagnet. The authors acknowledged a difficulty in obtaining a small moment with a large specific heat anomaly.

- Santini and Amato[39] suggested that the phase transition is an quadrupolar transition, (as in $UPd_3$) attributing the antiferromagnetism to a second transition induced by the presence of quadrupolar order.

- Gor'kov discussed the view that this phase transition involves a single hidden order parameter that violates time reversal symmetry, coupled weakly to the staggered magnetization.[40] Gor'kov suggested the intriguing possibility that the order is *spinorial* in character, associated with the development of spin anisotropy in the hybridization valence fluctuations of the atomic f-states.

Another aspect of the antiferromagnetism that has deservedly attracted attention, is its interplay with superconductivity. Here we appear to see rather contrasting behavior between the uranium HF superconductors, which happily co-exist with homogenious antiferromagnetism at the f-sites,[41] and $CeCu_2Si_2$, where the development of superconductivity appears to lead to the expulsion of antiferromagnetism.[42]

My personal view, is that it is hard to address the co-existence of antiferromagnetism and superconductivity without confronting the basic question of how the f-electron spins are involved in the superconducting order parameter. The pragmatic viewpoint, that one sheet of the Fermi surface undergoes a spin-density wave transition, leaving behind a residual Fermi surface to undergo a pairing transition tacitly avoids the real-space[43] aspect of the problem: each uranium spin is simultaneously involved in a magnetic and a superconducting condensation phenomenon. Let me pose a question that I return to in the next section:

• *Heavy fermion superconductivity is a spin-correlating process in which a large amount of local-moment entropy is quenched as part of the condensation process. How does the microscopic order parameter for the superconductivity involve the local moment spin degrees of freedom?*

Before leaving the topic of magnetism, I should like to draw attention to an aspect of heavy fermion magnetism that has, to date received scant attention. Suzuki et al. presented some lovely experimental data on the low carrier density compound $CeP$,[44] where a sequence of magnetic phase transitions develop in a high field at regularly space intervals in $1/B$. This result suggests an interesting interplay between Landau level filling-factors and antiferromagnetism that is crying out for theory.

## 5. Heavy Fermion Superconductivity

### (a) Order Parameter Symmetry

The issue of order parameter symmetry in HF superconductors, particularly $UPt_3$, continues to be a subject of diverse theoretical opinion.[45,46] Two rival Landau Ginzburg theories compete to explain the three phases in superconducting $UPt_3$. In both scenarios the gap function contains two complex order parameters :

$$\Delta_{\vec{k}} = \eta_a \phi_a(\vec{k}) + \eta_b \phi(\vec{k}).$$

The basic point of contention:

• *Do the two order parameters $\eta_a$ and $\eta_b$ belong to a two dimensional representation of the*

*crystal point group,[47,48] or is there an accidental degeneracy between the transition temperatures of two different one-dimensional representations, resulting in the co-existence of two order parameters with different symmetries ?[49]*

Regrettably, both scenarios depend on certain "accidental" features in the Landau Ginzburg theory: to explain these features an appeal to microscopic considerations is required. Within a two-dimensional representation, the observed multicritical point in the phase diagram occurs when off-diagonal gradient terms are fine tuned to zero. Sauls discussed how, within a weak coupling BCS model, the weak hexagonal anisotropy of the upper critical field in $UPt_3$ was consistent with these almost vanishing couplings.[45] In the picture with an accidental degeneracy of transition temperature, there is as yet no compelling origin for the "degeneracy". The worry:

- *Is it possible that the "accidents" appearing in the Landau Ginzburg models of superconducting $UPt_3$ reflect a hitherto unidentified symmetry of the order parameter?*

### (b) Quasiparticles, and the character of the condensation process.

BCS theory is such a beautiful and successful theory that we are naturally reticent to contemplate any departure from this framework. On the other-hand, the Cooper instability is so fundamentally tied to the presence of a Fermi surface that in a climate where Fermi liquid aspects of the normal state are under examination, it does not seem unreasonable to subject our most basic assumption, that HFSC involves a Cooper instability of a heavy electron Fermi surface, to similar scrutiny.

Within the family of six known heavy fermion superconductors, $UBe_{13}$ is a particularly troublesome candidate for a Cooper pairing scenario. Historically, superconductivity in $UBe_{13}$ was observed shortly after superfluidity in liquid $He-3$.[50] At the time, the likelyhood of a dense local moment system exhibiting superconductivity appeared so remote that the results were dismissed as an effect of filamentary impurities of pure $U$. Twenty years later, the established superconductivity of this remarkable metal still defies a consistent theoretical description. Recall that for $UBe_{13}$[51]

- ⋄ The condensation entropy $\Delta S \sim \gamma(T_c) \sim 1 J/mol/K$ is a significant fraction of $Rln2$ per mole, so the transition temperature is a large fraction of the Fermi temperature. No other superconductor has such a large condensation entropy. This is spin entropy, indicating that the processes of condensation and spin quenching of the local $U$ moments are <u>intimately related</u>.

- ⋄ The resistivity at $T_c$ is about $200\mu\Omega cm$, close to the Mott limit for this system; this means that electron quasiparticle scattering rates are comparable with the Fermi temperature. The heavy Fermi surface is presumably smeared beyond recognition.

- ⋄ The specific heat anomaly, $\Delta C_v/C_v|_n \sim 2.5$ can become enormous. Here in Amsterdam, Ahrens et al. reported that a tiny amount of boron doping actually doubles the specific heat anomaly to $\Delta C_V/C_V \sim 4$.[52]

To obtain such large ratios of $T_c/T_F$ and $\Delta C_v/C_V$ within an Eliashberg approach would require a huge coupling constant; inelastic scattering responsible for the $200\mu\Omega$ resistivity is severely pair-breaking, further exacerbating the plight of any proposed Cooper instability. In a conventional superconductor, the strong coupling limit corresponds to a condensation of pre-formed pairs:[53] in $UBe_{13}$, such objects would need to directly involve the localize moments. It would be an educational exercise to quantify these worries.

• *What parameters would be required in an Eliashberg description to explain the magnitude of $T_c/T_F$ and $\Delta C_V$ in $UBe_{13}$ in the presence of such large inelastic scattering?*

A more general, but equally disturbing quirk of HF superconductivity, is the failure of the various power-laws associated with the thermodynamics and quasiparticle response functions to fit consistently into a picture of anisotropically paired BCS superconductivity. As Heffner remarked at this meeting, "time and time again", the NMR and NQR relaxation rates in highly correlated superconductors, follow a $T^3$ dependence $1/T_1 \propto T^3$. This feature is traditionally interpreted in terms of a d-wave, or p-wave state with lines of gap zeroes. Crudely speaking, the NMR relaxation rate is given by a thermal average of the quasiparticle density of states $N(\omega)$ with the quasiparticle spin matrix elements $\langle \omega | S_\pm | \omega \rangle$ (spin coherence factors):

$$\frac{1}{T_1 T} \propto \int d\omega \left( -\frac{\partial f}{\partial \omega} \right) [N(\omega)]^2 \left| \langle \omega | S_\pm | \omega \rangle \right|^2$$

Here, the thermal average selects energies $\omega \sim k_B T$, so a simple rule-of-thumb is that the temperature dependence of the NMR relaxation rate depends on the square of the density of states and spin coherence factor at energy $\omega = \pi k_B T$:

$$\frac{1}{T_1} \sim T \left[ N(\omega) | \langle \omega | S_\pm | \omega \rangle | \right]^2_{\omega = \pi k_B T}$$

Within the framework of BCS superconductivity, spin coherence factors are unity at low energies. A $T^3$ dependence in $1/T_1$ thus indicates a linear dependence of the density of states $N(\omega) \propto \omega$, which in turn implies line zeroes of the gap. By contrast, the quasiparticle specific heat is given by

$$C_V \sim T \left[ N(\omega) \right]_{\omega = \pi k_B T}$$

So for a BCS superconductor, universal $T^3$ NMR relaxation rates should be universally accompanied by $T^2$ specific heats. Paradoxically, the universality of the NMR and NQR relaxation rate is not reflected in the specific heat capacities. This point is dramatically brought home by the latest NQR data of Kitaoka et al. on superconducting $UPd_2Al_3$.[54] The specific heat has the form $C_v = \gamma T + BT^3$ in the superconducting state.[43] Despite the absence of any measurable quadratic term in the specific heat, recent results of Kitaoka et al. show *four decades* of $T^3$ dependence in the zero field NQR relaxation rate. Why?

Unlike the early days of HF superconductivity, sample quality has reached a point where persistent qualitative discrepancies of this kind can not be blithely waived away. The failure of a single density of states to reconcile the NMR/NQR relaxation suggests that the spin matrix elements must have a non-trivial energy dependence. Phenomenologically, we may attempt to reconcile the $T^3$ NMR and NQR relaxation rates with a linear specific heat capacity, by hypothesizing a finite density of states of gapless quasiparticles whose spin coherence factors vanish linearly with energy.

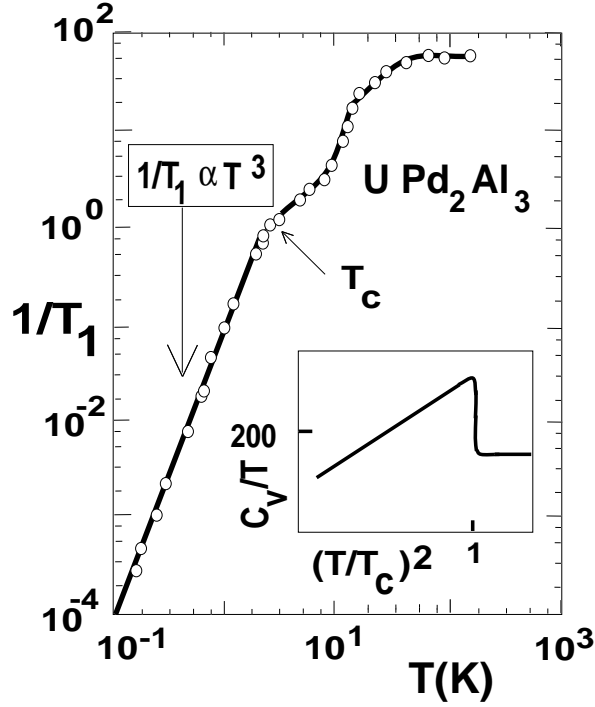

Fig. 3: **NMR/NQR paradox.** Illustrating the $T^3$ dependence of the NQR relaxation rate over *four decades* of relaxation rate in the superconducting phase of $UPd_2Al_3$.[54] Inset: a *cubic*, rather than quadratic temperature dependence of $C_v$.[43]

$$\left.\begin{array}{c} N(\omega) \sim \text{constant} \\ \\ \left|\langle\omega|S_\pm|\omega\rangle\right| \sim \omega \end{array}\right\} \quad \frac{1}{T_1} \propto T^3 \qquad (3)$$

Inside a superconductor the Boguilubov quasiparticles take the form

$$a^\dagger_{\vec{k}} = \underline{u}_{\vec{k}} c^\dagger_{\vec{k}} + \underline{v}_{\vec{k}}(i\sigma_2) c_{-\vec{k}}$$

Vanishing coherence factors occur when the magnitudes (eigenvalues) of $\underline{u}_{\vec{k}}$ and $\underline{v}_{\vec{k}}$ become equal. This occurs when the ratio of electron kinetic energy to pairing energy vanishes

$$\left(\frac{\text{kinetic energy}}{\text{pairing energy}}\right) = \left(\frac{\epsilon_{\vec{k}}}{\Delta_{\vec{k}}}\right) \to 0$$

As we approach a gap node in a BCS superconductor, this ratio goes to infinity rather than zero. The corresponding quasiparticles are <u>unpaired</u> and their coherence factors are unity. Vanishing coherence factors thus require a new type of theory for the condensed state with gapless paired quasiparticles.

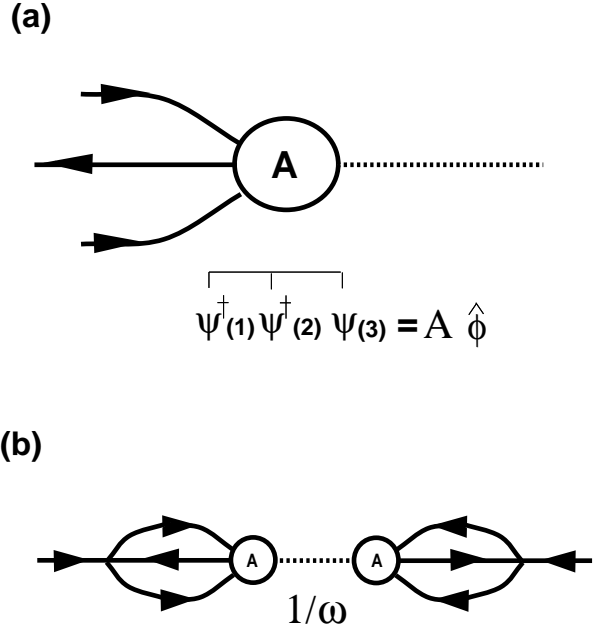

Fig. 4: **Odd frequency pairing.** Illustrating (a) anomalous amplitude for three particle bound-state and (b) anomalous pairing amplitude that scales as $1/\omega$ in this picture.[57,58]

A conceivable microscopic route to such vanishing coherence factors is the development of odd-frequency pairing. The possibility of a gap function that is an odd function of frequency $\Delta(\omega) = -\Delta(-\omega)$ dates back to early work by Berezinskii[55] and has received more recent attention from Balatsky, Abrahams, Schrieffer and Allen.[56] My work on this topic in collaboration with Miranda & Tsvelik[57,58] suggests that a very specific type of odd-frequency gap function might be driven by the very engine responsible for HF physics: the coherent Kondo effect. In a toy model, we found condensation into the superfluid state can occur with a coherent development of *three-body bound states*, driven microscopically by the binding of spins to conduction electrons.[57]

In this picture resonant scattering off the three-body resonance creates a triplet gap function that diverges at low frequencies as

$$\Delta(\omega) \propto \frac{1}{\omega}.$$

(Fig. 4.) Surfaces of gapless quasiparticles with vanishing spin and charge coherence factors develop in the condensate, predicting a thermal conductivity that remains finite in all directions as $T \to 0$. An attractive feature of this kind of process, is the appearance of a *staggered* phase. Cox et al. have recently demonstrated that in a hexagonal lattice,[59] translational invariance forbids gradient terms between different components of a staggered triplet order parameter, providing a fascinating mechanism for the symmetry-stabilization of the multicritical point in $UPt_3$.

## 6. Concluding remarks & experimental wish list

I should like to end this summary with a brief wish list of experiments that would shed light on some of the discussion we have had: success with any one of them would be already far more than a theorist has a right to expect.

- NMR and NQR at the f-site of a HF superconductor. Unlike the cuprate superconductors, we have no information about the nature of the spin fluctuations at the most active site. An NMR or NQR experiment on U-235 nucleii in an isotopically enriched sample of HF superconductor would provide invaluable information about the way the f-electrons are involved in the condensation process.

- dHvA oscillations for orbits in the basal plane of $UPt_3$. These orbits have not yet been resolved in *dHvA* experiments. Were this possible, one would be able to directly confirm the presence of a line of gap zeroes by following these oscillations into the superconducting state.

- Thermal conductivity at low temperatures in $UPt_3$. For a hybrid gap with line and point nodes, the anisotropy of the thermal conductivity in $UPt_3$ is predicted to diverge in good samples at low temperatures. Can this be observed?[60]

- High field experiments on the single plane cuprate superconductor 2201. If the zero mode driving NFL behavior is local in character, one might expect that the application of a high field would develop a finite Fermi temperature ($T_F \propto H^2$), producing a cross-over to a quadratic temperature dependence of the resistivity. In the single layer, cuprates with $T_c \sim 7K$ and low upper critical fields, it may be possible to observe the cross-over at low temperature and high fields.

Finally, please note that by stressing questions over answers, I really do not intend to suggest that one must be "deconstructionist" in developing new theory. The act of going back over fundamental questions is a rather essential process of renewal that leads to new ideas, but also strengthens our conviction and insight into aspects of previous theories and models that <u>are</u> right. For example, despite our interest in NFL behavior, there is as yet no compelling reason to abandon the Landau Fermi liquid as a low temperature description of the normal states of HF metals. Actually, this conference has been very successful in reaffirming a vital old consensus: that this is a really friendly and exciting field of condensed matter physics.

This work was supported by NSF grant DMR-93-12138. I should like to thank my colleagues at Rutgers University for their comments in reading this manuscript.

[32] H. von Löhneysen, *Non-Fermi liquid behavior in heavy fermion alloys"*, Th-I3.

[33] F. G. Aliev et al., *"Anomalous ground state in $U_{0.9}Th_{0.1}Be_{13}$"*, Th-P19.

[34] M. C. Andrade et al., *" Kondo effect and non-Fermi liquid behavior in the $U_{1-x}Th_xPd_2Al_3$ system"*, Th-P20.

[35] K. Ueda and T. Moriya, *"Spin Fluctuation Mechanism for the High Temperature superconductivity"*, Tu P-08.

[36] F. P. Onufrieva, J. Rossat Mignod and V. P. Kushnir, *"Spin and hole dynamics of high $T_c$ cuprates ...."* Tu P-10.

[37] S. Suga, A. Okiji and N. Kawakami, *" Elementary excitations for multichannel Kondo model"*, Tu-P74.

[38] A. Sikkemma et al. , *"Small moments in heavy fermions: a model Hamiltonian"*, Tu-C5.

[39] P. Santini and G. Amoretti, *"Quadrupolar model of the 17.5K phase transition in $URu_2Si_2$"*, We-P30.

[40] L. P. Gor'kov and V. Barzykin, *"Non-Neél ordering in presence of strong anisotropy"*, We-C8.

[41] R. Konno, *"Theory of interplay between antiferromagnetism and superconductivity"*, W-P24.

[42] P. Thalmeier,*"Coexistence of HF superconductivity and anisotropic electron-hole pairing,* We-P20.

[43] R. Caspary et al., *Phys. Rev. Lett.* **71**, 2136 (1993).

[44] T. Suzuki et al., *"Anomalous physical properties of the low-carrier concentration states in f-electron systems"*, Mo-I4

[45] J. A. Sauls, *"Phenomenology of the order parameters in unconventional superconductors"*,
21

[60] Recent measurements show a weak development of anisotropy in the thermal conductivity down to $0.1K$, B. Lussier, B. Ellman and L. Taillefer, *"Thermal conductivity of superconducting $UPt_3$"*, We-P09. See also K. Behnia, L. Taillefer et al., *J. Low Temp. Phys.* **84**, 261 (1991).